               \font\sixbf=cmbx6      
\def\apj{{\it Ap. J. }}                                
\def\apjl{{\it Ap. J. Lett. }}                         
\def\apjs{{\it Ap. J. Supp. }}                         
\def\prl{{\it Phys. Rev. Lett. }}                      
\def\aap{{\it A\&A }}                         
\def\jetp{{\it Sov. Phys. JETP }}                      
\def\mnras{{\it MNRAS }}                               
\def\ssr{{\it Space Sci. Rev. }}                       
\def\teq#1{$\, #1\,$}                         

{\catcode`\@=11                                                  
\gdef\SchlangeUnter#1#2{\lower2pt\vbox{\baselineskip 0pt\lineskip0pt    
\ialign{$\m@th#1\hfil##\hfil$\crcr#2\crcr\sim\crcr}}}}           
\def\gtrsim{\mathrel{\mathpalette\SchlangeUnter>}}

\documentstyle[]{aipproc}

\begin{document}
\newcommand{\vol}[2]{$\,$\bf #1\rm , #2}                 
\vphantom{p}
\vskip -45pt
\centerline{\hfill To appear in the Summary-Rapporteur Volume 
  of the 26th International}
\centerline{\hfill Cosmic Ray Conference, ed. B.~L. Dingus
 (AIP, New York, 2000)}
\vskip 15pt

\title{Cosmic Ray Origin, Acceleration\\ and Propagation}

\author{Matthew G.~Baring$^{\dagger}$}
\address{Laboratory for High Energy Astrophysics, Code 661\\
NASA Goddard Space Flight Center, Greenbelt, MD 20771\\
{\it baring@lheavx.gsfc.nasa.gov}\\
$^{\dagger}$Universities Space Research Association}

\maketitle

\begin{abstract}
This paper summarizes highlights of the OG3.1, 3.2 and 3.3 sessions of
the XXVIth International Cosmic Ray Conference in Salt Lake City, which
were devoted to issues of origin/composition, acceleration and
propagation.
\end{abstract}

\section*{Introduction}
 \label{sec:intro}

A review of a collection of papers on cosmic ray origin, acceleration
and propagation is necessarily broad.  Historically, International
Cosmic Ray Conferences have separated the papers in these extensive
subjects for consideration by different Rapporteurs.  However, since
the Rome conference in 1995, a new precedent has been established with
a review of all these fields becoming the responsibility of one
individual.  This has perhaps been propelled by the burgeoning number
of astrophysics-related contributions to the meetings, and has reduced
the comprehensiveness possible in a Rapporteur's written summary.  This
tract represents my attempt to assemble a description of interesting
and new results presented at the Salt Lake City conference pertaining
to the origin, acceleration and propagation themes.  Space limitations
preclude completeness, and accordingly I ask for forbearance from
authors who feel their work is not given a sufficient exposure here.  I
also offer the standard disclaimer: that the views expressed here are
personal, and may not reflect the perspective of ``The Management,''
i.e. the contributing authors whose research has provided such a
rewarding experience for this Rapporteur.

The material I was asked to report upon can be grouped into five
categories: origin and composition, for which there were \teq{\approx
11} papers, propagation of ions and electrons (26 papers), acceleration
theory and astrophysical applications of acceleration models
(33 papers), and discussions of ultra-high energy cosmic rays (UHECRs,
12 papers), with about 5 papers falling into the miscellaneous pot.
These themes define the structure of this review, and there have been
varying degrees of advancement in these fields.  The citation scheme
adopted here identifies conference papers by OG 3.*.* designations, for
which the reader should refer to the proceedings volumes, either
hardcopy or on-line at {\tt
http://krusty.physics.utah.edu/\~{}icrc1999/proceedings.html}.

\section*{Origin and Composition}
 \label{sec:origin}

The power supply for the acceleration of galactic cosmic rays is
traditionally attributed to supernova remnants, yet there is much
debate as to the mass and type of the progenitor stars, the specific
nature of the circumstellar environment, and the galactic origin of the
material accelerated.  This discussion has spawned a field rich with
ideas, with diagnostics largely provided by cosmic ray primary
compositional data.  The papers presented at this meeting generally
relate to one of two problems:  (i) the discussion of whether fresh
supernova ejecta or environmental dust grains provide the seeds
for cosmic ray acceleration, and (ii) explanations of the Li, Be, and B
abundances, the well-known LiBeB problem.

Early ideas on cosmic rays focused on the environment for their
acceleration, assuming some pre-existing seed population, rather than
addressing the question of the origin of such seed material.  Over a
period of time, it became clear that the galactic cosmic ray
(GCR)/solar photosphere abundance ratios provided valuable clues to the
origin of galactic cosmic ray matter.  Such ratios exhibit (e.g. see
Silberberg, Tsao \& Barghouty OG 3.1.06) a general enrichment of
refractory elements (i.e those with high condensation temperatures: Mg,
Al, Si, Fe and Ni) relative to highly volatile ones (principally H, He,
N, Ne and Ar).  Two competing interpretations of this property
emerged.  The first is that low energy ions are pre-accelerated in
stellar coronae to enrich the interstellar medium (ISM) before
participating in acceleration at proximate SNR shells.  In early work,
\cite{CasseGoret78,Meyer85} suggested that enrichment correlates with
elemental first ionization potential (FIP; see also Silberberg, et al.
OG 3.1.06), with high-FIP elements being somewhat suppressed.  The FIP
interpretation was largely driven by the discovery that FIP biases the
composition of solar energetic particles; hence the connection to
stellar coronae was made.  The second proposal originated with Bibring
\& Cesarsky \cite{bibces81} and Epstein \cite{Epstein80}, where erosion
products of grains formed from old material seed the acceleration
process, so that enrichment should correlate with volatility
\cite{mde97}.  The acceleration process is then naturally enhanced in
non-linear SNR shocks with increasing mass-to-charge (A/Q) ratio of the
species \cite{edm97} in a manner commensurate with observed
abundances.

While FIP proponents are invoking atomic physics concepts and a
volatility interpretation appeals to molecular physics, the two views
are not entirely opposite:  FIP and volatility are clearly related
quantities, albeit in a rather subtle manner.  For many light and
heavier sub-Fe elements, these two scenarios provide comparable
GCR/solar abundance ratios.  Yet success of the FIP-based models is
contingent upon a number of disconnected and controversial assumptions,
pertaining mostly to H, He, and $^{22}$Ne and the contribution of
Wolf-Rayet winds.  In contrast, the volatility description offers a
more coherent picture with fewer debatable assumptions, depending
principally on the chemistry and composition of interstellar grains.
It is therefore becoming the more widely-accepted description, with the
work of Lingenfelter \& Ramaty (OG 3.1.05) coming out in support of
volatility as the descriptor of cosmic ray abundances.  Nevertheless,
their research group had previously advocated \cite{lrk98} fresh
supernova ejecta as the seeds for acceleration, as opposed to the
grains created from older matter in the model of Meyer, Drury \&
Ellison \cite{mde97,edm97}.  This was a major point of controversy that
was addressed and resolved at the Conference, based on two
discriminating pieces of information.

The first diagnostic concerned the C and O ratios.  These elements
provide critical diagnostics since they possess intermediate FIPs and
are moderately volatile, and hence are bridge elements between the
volatiles and the refractories.  Both are key products of
nucleosynthesis in massive O and B stars, which are the progenitors of
the type Ib and II supernova that dominate the observed supernova
population.  The property crucial to the success of the
grain-acceleration proposition is that these two species are present in
grains (e.g.  various oxides and graphite) in just the appropriate
amounts to explain their abundances \cite{mde97}.  Consequently, it
becomes apparent that interstellar grain chemistry is the important
parameter for the composition problem, and should be a focus of future
research efforts.

The second decisive indicator concerned the age of the seeds for
acceleration.  Since grains can be much older than the SNRs that tap
them, the grain-induced cosmic ray composition picture
\cite{mde97,edm97} is less subject to temporal restrictions provided by
unstable nuclei that offer markers of the chronology of
nucleosynthesis.  Foremost among these is the electron K-capture decay
of $^{59}$Ni to $^{59}$Co, with a half-life of around \teq{10^5} years,
for which the ACE experiment has recently provided discriminating
information: the low abundance of $^{59}$Ni relative to Fe and the high
abundance of $^{59}$Co (Weidenbeck et al., OG 1.1.01) implies a passing
of at least \teq{10^5} years between nucleosynthesis and acceleration.
Meyer, Drury \& Ellison have consistently argued that grains are easily
old enough to satisfy the ACE temporal constraints.  While Lingenfelter
et al. \cite{lrk98} had advocated a fresh ejecta scenario, Higdon,
Lingenfelter \& Ramaty's contribution (OG 3.1.04) indicated an
evolution in their position so that the two groups concurred that the
ACE dataset does indeed provide age lower bounds that render fresh
(i.e. young) ejecta unlikely seeds for the acceleration process at SNR
shocks.  Focus has now turned to timescales of ejecta mixing well in
excess of \teq{10^6} years, which can be suitably probed (Waddington,
OG 3.2.33) by abundance measurements of actinides such as Th, Np, CM
and Pu.  Westphal (OG 3.1.09) discussed the potential for the ECCO
experiment aboard the International Space Station to provide such
discriminating data.

The mixing question is pertinent to the discussion of whether
superbubbles with many SNIb/SNII explosions as opposed to more isolated
ISM regions with SNIa progenitors are the locales for cosmic ray
origin.  The site issue, still unresolved, provides a natural
progression to the LiBeB problem.  This longstanding conundrum relates
to the abundances of Li, Be and B in old halo stars, the principal
spallation products of reactions of nucleosynthetic or ambient
$^{12}$C, $^{14}$N and $^{16}$O in collisions with hydrogen and helium
of either ISM or ejecta origin (see, e.g. Korejwo et al. OG 3.2.22, for
accelerator data on $^{12}$C fragmentation/spallation cross-sections
for various products in the GeV/nucleon range).  Balmer-like line (Be
II) observations indicate a linear correlation (e.g. Ramaty,
Lingenfelter \& Kozlovsky, OG 3.1.03; Fields \& Olive OG 3.2.04) of the
abundance of LiBeB with Fe metallicity (Fe/H) in these metal-poor stars
(note that Fe/H is effectively an age parameter for these systems).
Yet, theoretically (see the review in \cite{vroc98}) LiBeB is expected
to increase quadratically with Fe/H, since, for a constant supernova
rate, LiBeB/H should scale as the integral over time of the supernova
rate times the total number of antecedent supernovae in the galaxy.
This apparent conflict becomes cleaner, observationally, by considering
Be alone, since it provides no ambiguities; some of the $^7$Li is
probably a product of primordial nucleosynthesis, and much of the
$^{11}$B population may result from neutrino-induced spallation (on
$^{12}$C) in supernovae.

Presentations on explaining Be/H evolution at the Conference included
papers by Ramaty, Lingenfelter \& Kozlovsky (OG 3.1.03) and Parizot \&
Drury (OG 3.1.18, OG 3.2.51).  A result common to these two groups is
that, by separating the light and metallic spallation participants in
space, a decoupling between the metallicity of stars and their age is
effected.  This is achieved if there is no significant mixing between
metal-poor ISM that is accelerated at a supernova remnant's forward
shock and the enriched, high-metallicity ejecta accelerated at a
remnant's reverse shock (Parizot \& Drury OG 3.1.18).  The dominant
contributions to Be production are then spawned by (i) low metallicity
ISM ions accelerated by forward shocks colliding with metal-rich
supernova ejecta, and (ii) enriched ejecta material accelerated at
reverse shocks interacting with light elements from the surrounding
ISM.  In each case, the Be production is independent of the ISM
metallicity, generating a Be/H halo star abundance proportional to Fe/H
metallicity.  For this reason, Ramaty, Lingenfelter \& Kozlovsky (OG
3.1.03) argue that the Be/H evolution with Fe/H is a strong indication
that fresh ejecta are crucial to cosmic ray origin.  Reconciling the Be
production with the ACE observations of $^{59}$Ni should be a major
objective of future studies.  Supernova/cosmic ray energetics also play
a constraining role in this discussion, with both Ramaty, et al. and
Parizot \& Drury observing that there is an underproduction of Be (by
over an order of magnitude) in the early galaxy if most supernovae
explode in the average ISM.  This has motivated papers by Higdon,
Ramaty \& Lingenfelter (OG 3.1.04) and Parizot \& Drury (OG 3.2.51)
that describe how a superbubble/starburst locale for LiBeB generation
can provide prepared metallicity-enhanced environs due to the OB
stellar associations.  The production rate can increase more than
tenfold to match the observed abundances in this scenario because the
spallation reactions involving enriched ambient CNO can tap the greater
accelerating potential of forward shocks in SNRs.

A cautionary note for the LiBeB problem was sounded by Fields and Olive
(OG 3.2.04).  While historically Fe/H has been used as the marker of
metallicity for discussing Be production, Fields and Olive argued that
the O/H ratio is a far more appropriate indicator since oxygen is an
actual participant in the spallation reactions that spawn Be.  The
consequences of such a shift in perspective are substantial.  The
evolution of O/H does not trace Fe/H linearly so that O and Fe are
different indicators of metallicity.  Accordingly, Fields and Olive
observe that Be/H is more strongly dependent on O/H than Fe/H in
halo (population II) star atmospheres, more closely
resembling the quadratic dependence that was anticipated in incipient
theoretical considerations of the LiBeB abundances.  The implication of
their work is that the so-called LiBeB problem is a ``tempest in a
teapot.''  This is not entirely discouraging for theorists in their
quest for nailing the origin of cosmic rays, since the data spread in
the Be/H versus O/H diagram is considerable; observationally, it is more
problematic to determine oxygen metallicity than Fe/H.  Future refined
observations from uniform/consistent stellar atmospheres should resolve
this issue.

\section*{Propagation}
 \label{sec:propagation}

Studies of propagation have perhaps had the slowest evolution of the
sub-fields covered here.  This is essentially imposed by the pace at
which new and discriminating experimental data relating to this complex
problem are forthcoming.  Properties of the interstellar medium of the
galaxy remain enigmatic, presently prohibiting the elimination of any
one of the handful of preferred propagation models.  Foremost in this
group is the ``canonical'' Leaky-Box approximation, the ``tool of
choice'' for most members of the propagation community, due to its
simplicity.  More sophisticated  and physically realistic models with
various mutations are the halo diffusion picture, wind
scenarios, turbulent diffusion model, and calculations invoking
re-acceleration, each with their proponents (see \cite{berez90} for a
review).  There are a number of standard tests for the viability of
each of these; we shall explore here the latest results separately for
the cases of propagation of ions and electrons.

\subsection*{Ions}
 \label{sec:prop_ions}

A significant number of papers were presented, many producing very
similar results.  The leaky box model (LBM), where a ``one-zone''
scenario is envisaged with an escape length or rather grammage
\teq{X_{\rm lb}} forming the principal model parameter, and the halo
diffusion picture (HDM), where the galactic disk and halo represent two
regions distinct in their source and diffusion properties, were the
most common invocations (e.g. OG 3.1.16, 3.2.02, 3.2.03, 3.2.06,
3.2.07, 3.2.08, 3.2.09, 3.2.18, 3.2.32).  While these two models
dominate the discussion here, propagation in galactic winds (OG 3.1.16,
3.2.07, 3.2.13, 3.2.19, 3.2.32) and contributions from re-acceleration
(OG 3.2.02, 3.2.07, 3.2.18, 3.2.32) were also considered.

In the LBM, the grammage parameter is often specified as a
broken-power-law in rigidity \teq{R} (e.g. Ptuskin et al. OG 3.2.02),
increasing as a moderate power of particle velocity \teq{\beta} at
non-relativistic speeds and declining roughly as \teq{X_{\rm lb}\propto
R^{-0.6}} for relativistic energies \teq{E}.  This form is chosen (i)
to explain the observed steepening of the primary cosmic ray spectrum
from the approximately \teq{E^{-2.1}} spectrum expected at sources,
(ii) match the observed secondary/primary ratios of stable species, and
(iii) to accommodate spectral shapes observed in the transition region
between the modulated and unmodulated ion spectrum.  Coefficients of
these proportionalities are of the order of a few g/cm$^2$ to match
densities of the interstellar medium and establish scale-heights above
the galactic plane of the order of a kpc or so.  Physically, the decline
in \teq{X_{\rm lb}} as a function of rigidity corresponds to the
expectation of greater losses for more energetic particles.  The cosmic
ray production in the LBM is homogeneous in space, not being coupled to
the galactic plane.

The halo diffusion model (e.g. \cite{ptuskin74}) introduces more
complexity, distinguishing between galactic disk and halo with
different source densities and propagation characteristics in each
region.  Spatial uniformity can be assumed in each region (e.g.
Ptuskin et al., OG 3.2.02, OG 3.2.32) or disk and
halo can possess inhomogeneous distributions in altitude \teq{z} above
the plane (e.g.  Strong \& Moskalenko, OG 3.2.18).  The diffusive
escape parameter is usually set to \teq{X_{\rm e}\propto 1/{\cal
D}\propto R^{-1/3}} in accord with the dependence of the diffusion
coefficient \teq{{\cal D}} for Kolmogorov turbulence.  Essentially,
free escape arises at the halo extremities in this scenario, and the
selective confinement of matter near the plane renders the pathlength
distribution for losses exponential as in the Leaky Box model.  The
vertical height of the disk is constrained by the diffusive lengthscale
\teq{\sqrt{{\cal D} \tau}} for ``interesting'' radioactive isotopes of
ballistic lifetime \teq{\tau} (i.e. \teq{\sim 10^6} years; discussed
below).  A distinct advantage of the HDM is that it can accommodate the
observed low cosmic ray anisotropies that are almost constant out to
\teq{10^{14}}eV (e.g. \cite{kifune91}; see also Hillas OG 3.2.10) more
easily than the LBM, due to its weaker dependence of loss scale on
rigidity.

Primary source spectra for species such as carbon and iron alone are
insufficient to discriminate between Leaky Box and halo diffusion
models (e.g. see Ptuskin et al. OG 3.2.32), being more dependent on
solar modulation properties (such as the assumed force-field
potential:  e.g. Webber, OG 3.2.8, Strong \& Moskalenko OG 3.2.18).
Stable secondary to primary ratios are somewhat more sensitive to model
characteristics since they probe energy loss rates in matter traversal,
i.e. \teq{X_{\rm lb}} and \teq{X_{\rm e}}, for different species
involved in nuclear interactions with the interstellar medium.  The
most popular choices for these ratios, corresponding to spallation
reactions involving the principal components of cosmic rays, are those
of boron to carbon, B/C, and sub-iron group to iron nuclei,
(Sc+Ti+V)/Fe.  The spectrum of the spallation products traces that of
the parent nuclei when they are created, with a subsequent steepening
being induced by the energy-dependent propagation effects.
Nevertheless, the increased data spread appearing in such ratios is
sufficient to preclude unequivocal discrimination between models, so
that the LBM and HDM are equally viable (e.g. OG 3.2.32) based on
analysis of stable secondaries.

Hence considerable effort was expended in a number of papers that
focused on radioactive isotopes.  The abundances of suitable secondary
radioactive nuclei provide clues to the confinement time of cosmic rays
in the galaxy (e.g. Streitmatter \& Stephens OG 3.2.03), and therefore
offer observational diagnostics complementary to those engendered by
matter traversal.  Suitability is naturally governed by significant
elemental abundances and lifetimes that approximate typical galactic
disk diffusion timescales of 1 Myr.  Therefore, excellent choices
include $^{10}$Be (beta decay, 2.3 Myr), $^{26}$Al (inverse beta decay/
K-capture, 1.6 Myr) and $^{36}$Cl (beta decay, 0.4 Myr); $^{54}$Mn is
also a possible option, though its \teq{\beta^+} decay lifetime is
still not precisely determined.  While often-quoted Al/Mg and Cl/Ar
fractions represent parent/daughter nuclei pairs, the ratio of choice
for $^{10}$Be decay is $^{10}$Be/$^{9}$Be, representing the relative
abundance of surviving $^{10}$Be to its ``sister'' spallation product
$^{9}$Be rather than its decay offspring $^{10}$B.  This alternative is
afforded by the well-measured cross sections for spallation reactions
in accelerators.  Note that $^{10}$Be is optimal for experimental
purposes due to the lower mass resolution required to distinguish it
from other isotopes.  Of particular interest is the trans-relativistic
regime of 1--10 GeV/nucleon, where time-dilation effects are sampled.

Since the mean proximity of sources from the solar system differs for
the Leaky Box and halo diffusion models, the fractional abundances of
radioactive nuclides expected for the two scenarios are generally
disparate.  Various data model comparisons were presented by Ptuskin,
Soutoul \& Streitmatter (OG 3.2.02), Streitmatter \& Stephens (OG
3.2.03), and Simon \& Molnar (OG 3.2.06), sometimes expressed as
relative abundances (the experimentalists' preference), and sometimes
as surviving fractions (perhaps the theorist's choice), which
incorporate model-dependent information.  Variations in theoretical
predictions were modest, and preference for either the LBM or HDM is
indiscernible given that model parameters can be appropriately
fine-tuned; the abundance ratio data from Voyager, Ulysses and HEAO-3
missions are typically accurate to only a factor of two.  Yet the
potential for advances in this field in the near future is
significant.  The recent ACE data from the CRIS experiment (e.g.
Yanasek et al. OG 1.1.03, and Weidenbeck's highlight talk, these
proceedings) reduced experimental uncertainties in these ratios in the
0.1--0.3 GeV/nucleon range down to the 20\%--40\% level.  Further gains
are anticipated with ISOMAX (Hams et al., OG 3.1.33), which will extend
the range of exploration up to a few GeV, so as to more completely probe
the mildly-relativistic regime.

The possible influence of galactic winds and interstellar cosmic ray
re-acceleration complicate the propagation problem.  Winds away from
the galactic disk (typically at \teq{\gtrsim 20}km/sec) necessarily
enhance loss rates and therefore can impose less stringent requirements
on the energy dependence of the diffusion and lead to anisotropies in
the diffusion tensor (Breitschwerdt, Dogiel \& V\"olk, OG 3.2.19);
these authors argue that such winds may explain the small ratio of
radial gradients of diffuse gamma rays to cosmic rays.  Ptuskin et al.
(OG 3.2.32) indicate that wind and minimal re-acceleration models are
both just as consistent with stable secondary/primary ratio data as the
LBM and HDM.  Re-acceleration models did not achieve the same exposure
and topicality as in previous Cosmic Ray Conferences.  Their basic
properties are understood.  Depletions of low energy cosmic rays due to
{\it in transit} acceleration effectively eliminate the need for a
broken power-law for the variation of the escape length \teq{X_{\rm
lb}} with rigidity.  Re-acceleration alleviates the problem of weakly
rigidity-dependent, low-level anisotropies, by permitting a reduced
dependence of the escape length on \teq{R}.  At the same time,
re-acceleration has a profound influence on ions below 10 GeV/nucleon
(Jones et al. OG 3.2.07) that have long residence times; this becomes
an asset when trying to fit B/C and (sub-Fe)/Fe spectral flattenings in
the low-energy modulation range.

In concluding the discussion of ion propagation, note that two
formalism papers were contributed by Forman (OG 3.2.11) and Ragot
(OG 3.2.45), which focused on quasi-linear theory aspects of particle
diffusion in field turbulence (gyro-resonant and non-resonant,
respectively), works that while interesting for propagation
specialists, are more salient to heliospheric issues in the SH
sessions.

\subsection*{Electrons}
 \label{sec:prop_elec}

Considerations of electron propagation were largely confined to the
work of one research group, Webber and his collaborators.  Nothing
extremely new was forthcoming, yet discussion of electrons provides an
interesting forum for the interplay between cosmic ray physics and
astrophysics.  The observed cosmic ray (total) electron spectrum is
steeper in the 3--100 GeV range than its ion counterpart
\cite{mueller95}, suggesting either that ions and electrons possess
distinct propagation characteristics, or that electron source spectra
are steeper than ion ones.  This latter alternative was promoted in
several papers: Stephens (OG 3.2.14), Higbie et al. (OG 3.2.15),
Rockstroh et al.  (OG 3.2.16) and Peterson et al. (OG 3.2.17).
Inferences in this direction are facilitated by broadening the dynamic
range of cosmic ray energies sampled using data of astronomical
origin.  The diffuse radio synchrotron spectrum is very informative
since it evades modulation effects, and can therefore probe lower
electron energies, principally in the 0.2--3 GeV range.  However, the
``model-independence'' of such information is marred at low energies by
significant free-free absorption in the ISM (Peterson et al. OG
3.2.17).  Matching normalizations of the radio-derived \teq{e^-}
spectrum with the cosmic ray electron one measured at higher energies
requires assuming a mean interstellar field of around 5$\mu$G.  While
the aforementioned papers advocated an \teq{E^{-2.4}} electron source
spectrum, the data spread is sufficient to render an \teq{E^{-2.25}}
spectrum not implausible for the particular diffusion model invoked by
Rockstroh et al and Peterson. et al.  Since deductions pertaining to
the cosmic ray origin are contingent upon propagation and modulation
assumptions, the flatter source spectra are not presently excluded.

Higbie et al.  (OG 3.2.15) argued that modelling the diffuse gamma-ray
emission with the same Monte Carlo propagation simulation again points
towards a steeper \teq{e^-} source distribution: simultaneous fitting
of the pion ``decay bump'' in the $>50$ MeV EGRET data and the
relatively steep COMPTEL 1--30 MeV spectrum with a bremsstrahlung
component \cite{Strong93} (both experiments were on board the Compton
Gamma-Ray Observatory) provides the basis for this assertion.  Porter
\& Protheroe (OG 3.2.38) arrive at a different conclusion when
modelling diffuse gamma-ray emission, arguing in favour of flatter
electron source spectra.  These disparate inferences largely reflect
differences in propagation models, and therefore indicate the limits
that should be placed on such assertions at this stage.  Stephens (OG
3.2.14) addressed positron propagation and claimed a small (10--15\%)
charge-sign dependence of modulation; while potentially interesting,
data uncertainties limit this interpretation to merely a prediction for
future experimental verification.


\section*{Acceleration Theory and Astrophysics}
 \label{sec:accelastro}

The subject area of the theory of particle acceleration and
astrophysical applications was the most diverse in terms of the
material presented at the Conference.  Hence, only principal focal
points can be addressed in this brief exposition.

\subsection*{Acceleration Theory}
 \label{sec:accel}

The discussions of cosmic ray propagation hinge on the widely-used
assumption that the sources of cosmic rays produce quasi-power-law
populations (\teq{E^{-\alpha}}) with \teq{\alpha\approx 2.1}--2.4.
This is readily satisfied by {\it test-particle} acceleration at the
strong shocks formed at supernova remnant shells as the expansion
ploughs through the ISM.  This feature has lead to the almost universal
acclaim that SNRs are the site of cosmic ray acceleration, at least up
to the knee at \teq{\sim 10^{15}}eV.  Yet there are many subtleties,
including those related to deviations from the test-particle
approximation, how shock heating of the downstream gas is influenced by
the fluid dynamics, questions of the efficiency of injection
(particularly for electrons), and what are the differences between
relativistic and non-relativistic shocks.

The issue of validity of the test particle approximation is
important for the cosmic ray problem.  The beauty of diffusive
acceleration was underscored by the natural explanation it provided for
the power-law slope of the cosmic ray distribution over many decades in
energy.  Yet this attractive feature is contingent upon two criteria:
(i) that the accelerated particles do not modify the dynamics of the
shocked flow, i.e. act only as test particles to the problem, and (ii)
that there is no particular energy scale for losses of particles.  It
is palpable that neither of these properties is satisfied in shocks in
SNR shells, thereby eliminating the most aesthetic reason for
considering shock acceleration as the principal means of energizing
cosmic rays.  Nevertheless such acceleration is virtually inevitable
at the interface between supersonic and subsonic flows, and hence is
widely accepted to be ubiquitous in astrophysical systems by theorists
and experimentalists alike.

Non-linear shock acceleration effects and their implications featured
prominently in the contributed papers, and are suitably discussed in
the reviews of \cite{Drury83,JE91}.  When the accelerated ions have
sufficient pressure to modify the flow dynamics in the shock environs,
they can no longer be considered as test particles.  The cosmic ray
ions act to slow down the flow upstream of the shock discontinuity,
resulting in an increase of the overall compression ratio \teq{r} above
the canonical test-particle value of \teq{r=4} if the system sustains
significant losses of particles or energy.  This strengthening of the
shock adds to the non-thermal ion pressure, modifying the flow speed
further, and thereby provides a feedback that defines the non-linearity
of the acceleration process.  Such non-linear effects are present in
SNR shocks because they are inherently strong, have had sufficient time
(at least in the Sedov phase) to accumulate significant pressure in the
cosmic rays, and suffer losses on the largest spatial scales.
Electrons seldom contribute to the dynamics (e.g. \cite{bergg99}),
unless they possess a peculiarly large abundance relative to the cosmic
ray \teq{e/p} ratio \cite{mueller95} of 1--3\% in the 1--10 GeV range.

A principal signature of these non-linearities in strong SNR shocks is
the upward spectral curvature \cite{eich84} in the non-thermal ions,
a consequence of higher energy ions generally having larger diffusive
scales and thereby sampling greater effective compression ratios in the
cosmic ray-modified flow.  Concomitantly, the acceleration is enhanced
with increasing mass to charge (A/Q) ratio, implying a relative
profusion of higher metallicity species that was salient for the
cosmic ray origin discussion above.  Berezhko \& Ksenofontov (OG
3.3.09) and Ellison et al. (OG 2.2.09) illustrate such predictions of
non-linear acceleration theory and emphasize that spectral curvature is
consistent with all-particle or individual species data given the
significant experimental spread below the knee, an argument supported
by Zatsepin \& Sokolskaya (OG 3.1.02).  This line of reasoning is
obviously at odds with the common wisdom that the cosmic ray spectrum
is a beautiful power-law.  Merit can be found in both perspectives,
which are not inherently incompatible: the spectral curvature predicted
is sufficiently small (enhancements by a factor of a few over several
decades in energy) that it essentially cannot be discriminated from
exact power-laws as an appropriate model for the cosmic ray spectrum
below the knee.  In any case, since the cosmic ray measurements
represent a convolution of source properties and propagation
characteristics, such a distinction loses meaning.  In this regard,
gamma-ray signatures in the GeV to TeV band from isolated remnants will
be more informative in seeking evidence of spectral curvature.  

The critical point for discussion is that spectral cutoffs expected in
SNRs (generally around 10--100 TeV; see \cite{bergg99}, Berezhko \&
V\"olk, OG 3.3.08, Yoshida \& Yanagita, OG 3.3.11) could impose
structure in the cosmic ray spectrum more severe than observed near the
knee.  This is a principal outstanding problem for cosmic ray studies;
its resolution requires more detailed spectral and compositional
information in the vicinity of the knee (the ACCESS project
\cite{access99} should help provide this).  The KASCADE air shower
experiment provided some interesting results salient to this issue,
namely deductions of proton and Fe spectra from muon data (Haungs et
al. HE 2.2.02; Chilingarian et al. HE 2.2.04).  Complementary
inferences from gamma-ray upper limits (CASA-MIA results: Markoff et
al. OG 3.3.18; HEGRA observations: Horns et al. OG 3.2.24) are
currently not constraining.

Non-linear acceleration-induced spectral curvature obviously will
impose more severe requirements on propagation models, both by
requiring a stronger dependence of the escape length on rigidity and by
increasing difficulties in minimizing anisotropies of the highest
energy particles in the galaxy.  Another non-linear feature is the
reduction of the compression ratio of the viscous subshock (i.e. shock
discontinuity) below \teq{r=4}, thereby reducing the dissipational
heating of the downstream plasma (Ellison \& Berezhko OG 3.3.12).  This
property is pertinent to the interpretation of X-ray line emission from
SNRs, computations of X-ray bremsstrahlung in SNR emission models and
the deduced electron-to-proton ratio \cite{ebb99}; the latter impacts
the gamma-ray flux expected from remnants\cite{bergg99}.  Several
papers were devoted to such astrophysical signatures and are discussed
below.

The injection issue was the subject of two papers, Gieseler, Jones \&
Kang (OG 3.3.20) and Sugiyama \& Fujimoto (OG 3.3.21), though neither
paper treated electron injection, a perennial concern for theorists.
Gieseler et al. developed Kang \& Jones' diffusion-convection equation
approach to modelling acceleration at non-linear shocks by
incorporating a description, due to Malkov, of the interaction of
thermal ions with self-generated magneto-hydrodynamic (MHD) waves.  It
is unclear what advantages this step has to offer over antecedent
developments by Kang \& Jones that parameterized injection efficiencies
(e.g.  see OG 3.3.32, which discussed an interesting use of an adaptive
mesh technique to improve the dynamic range of lengthscales that can be
probed).  The injection formalism incorporated in OG 3.3.20 is based on
quasi-linear theory, which has limited applicability to turbulence in
the environs of strong, modified shocks.  Sugiyama \& Fujimoto
simulated injection in such strong turbulence by computing ion motions
in large amplitude MHD waves, using techniques employed in hybrid and
full plasma simulations.  Their test particle investigation of
essentially coherent acceleration in time-dependent electric fields in
the shock neighbourhood yielded expected results, which are usually
generated by more complete plasma simulations (reviewed in
\cite{JE91}), namely that suprathermal ions are produced in significant
numbers on timescales considerably larger than the ion gyroperiod.
Such coherent effects are an integral part of the dissipational heating
in the shock layers, and naturally provide injection that seeds
diffusive acceleration at higher energies.

From a small {\it pot pourri} of papers treating diverse
acceleration problems, I wish to highlight two contributions before
proceeding to the astrophysically-oriented offerings.  The first was
the presentation of a simple analytic model of non-linear acceleration
in plane-parallel shocks by Ellison \& Berezhko (OG 3.3.12), specifying
a complete (and continuous) particle distribution via a thermal
component plus a three-piece broken power-law representing non-thermal
ions.  The power-law slopes, energies of connection between the various
spectral portions, and the normalization coefficients are
self-consistently determined in a modelling of the flow hydrodynamics;
only the efficiency of injection from thermal energies need be
specified as a parameter.  The model possesses great potential for
astrophysical applications, due to its facility, and agrees well with
more complete predictions of Monte Carlo \cite{ebj96} and
kinetic transport equation \cite{byk96} techniques.

The second interesting result was in the discussion by Drury et al. (OG
3.3.13, OG 3.3.16) of ``pile-ups'' in cosmic ray electron source
distributions near the maximum (i.e. cutoff) energy due to significant
synchrotron losses.  This issue has had various preceding treatments,
with the conclusion that only test-particle shocks with compression
ratios \teq{r>4} could yield a build up of electrons near the cooling
cutoff, i.e. an improbable occurrence.  The new feature of Drury et
al.'s work is that momentum-dependent diffusion scales are treated so
that synchrotron cooling of electrons sufficiently remote downstream from
the shock can result in losses from the system additional to those due to
convection.  The criterion for build-ups relaxes to \teq{r\gtrsim
3.5}, generating an interesting regime of phase space where strong
shocks potentially can yield these spectral bumps.  Essentially,
pile-ups arise when momentum losses in cooling outpace the spatial
losses that are integral in determining the index of the canonical
test-particle distribution.  Such pile-up considerations could
prove very relevant to the interpretation of non-thermal X-ray emission
and TeV gamma-ray spectra from SNRs.

\subsection*{Astrophysical Applications}
 \label{sec:astro}

Supernova remnants were the dominant subject of astrophysical
applications of acceleration theory.  While dynamical calculations of
cosmic ray acceleration at SNR shocks and limited models of radio to
gamma-ray emission from these particles have been around for a long
time, this field has really burgeoned in the last half decade following
the detection by the EGRET experiment on the Compton Gamma-Ray
Observatory of a number of unidentified 100 MeV--10 GeV gamma-ray
sources with SNR celestial associations \cite{espos96} and the
subsequent campaigns \cite{prosch96,buck97} by atmospheric \v{C}erenkov
telescopes to search for TeV emission from various prime candidate
remnants (see \cite{baring00} and Buckley's Rapporteur paper in these
proceedings for reviews of this field).  The field now possesses
confirmed detections in {\it non-thermal} X-rays and TeV gamma-rays in
a few sources, an enviable position compared with the status 5 years
ago.  The models have rapidly become more sophisticated and complete in
their radiation predictions.  Two alternative techniques are at the
forefront of this acceleration problem, both being represented at the
Conference: (i) Berezhko et al.'s semi-analytic solution \cite{byk96}
of the time-dependent spherical transport equation for ions,
and (ii) Ellison, Baring and collaborators' use of a Monte Carlo
simulation of diffusive acceleration \cite{bergg99,ebj96}.  These
approaches each have their virtues and limitations.  Berezkho et al.'s
method handles all the time-dependent effects self-consistently, but
requires a parametric specification of injection, whereas the Monte
Carlo simulation, which automatically injects ions from the thermal
populations, models steady-state parallel shocks and incorporates
effects of time-dependence through a hybridization \cite{bergg99}
involving Sedov evolution of shock parameters.  Both methods must
parameterize electron injection, an imposition due to current
shortcomings in acceleration theory.

There is a remarkable convergence of results from these two
complementary models, as is patently evident in the spectral comparison
presented by Ellison \& Berezhko (OG 3.3.27).  While there are some
fine-scale dissimilarities, this global agreement has led to a fairly
robust set of predictions \cite{ebb99} for radio, X-ray and gamma-ray
astronomy, embodied in the Conference papers of Berezhko and V\"olk (OG
3.3.08), Berezhko, Ksenofontov \& Petukhov (OG 3.3.23) and Ellison et
al. (OG 2.2.09).  Principal features include the virtual constancy (and
peaking) of the maximum particle energy and gamma-ray luminosity
throughout the Sedov epoch (OG 3.3.08, OG 3.3.23), and prominent pion
decay emission for high circumstellar densities in both the GeV and TeV
wavebands; for ambient fields approaching 1 mG, synchrotron cooling is
sufficient to render such hadronic emission dominant in the super-TeV
range (Ellison et al. OG 2.2.09, and Berezhko \& V\"olk OG 3.3.24, who
also explore remnant properties for explosions in wind bubbles spawned
by massive progenitors).  Such pion decay signatures are potentially
almost unambiguous evidence of the presence of cosmic rays in supernova
remnants.  The quest for such a proof of cosmic ray acceleration in
SNRs is of primal importance to the cosmic ray community.  Acquisition
of this evidence seems imminent, given the impending ground-based and
spaced-based gamma-ray experiments scheduled to come ``on-line'' in the
next 5--6 years.  Theory is currently well-placed to interpret the
anticipated wealth of new information to be afforded by these
programs.

There was a marked paucity of papers addressing relativistic shocks at
the Conference.  This was in spite of considerable recent interest in
their acceleration properties by modellers of the topical gamma-ray
burst (GRB) phenomenon, and the probable relevance to generation of
ultra-high energy cosmic rays.  Baring (OG 2.3.03) provided the
principal offering at the Conference on acceleration predictions at
relativistic shocks, highlighting the major needs for GRB theorists:
quantifying the injection efficiency (particularly for electrons), and
determining the spectral index (which is not uniquely specified in
terms of the shock compression ratio) and the time and maximum energy
of acceleration.  None of these properties can be discerned easily, and
there is a major need to redress such gaps in our knowledge.  Baring
explored spectral differences between large angle scattering and pitch
angle diffusion in ultrarelativistic plane-parallel shocks (i.e. those
with bulk Lorentz factor \teq{\Gamma\gg 1}), and confirmed the finding
of Bednarz \& Ostrowski \cite{bo98} that in the case of pitch angle
diffusion, the power-law spectrum for accelerated particles approaches
approximately \teq{E^{-2.2}} as the shock speed asymptotes to the speed
of light.  Ostrowski (OG 3.3.07) discussed the possibility of
acceleration at shear layers bordering relativistic jets in active
galaxies.  As intuitively expected, he observed the acceleration to be
rapid due to large kinematic boosts acquired when particles diffuse
between the jet and surrounding medium.  Yet no indication of the
efficiency of injection was proffered, and it is unclear that this type
of boundary layer acceleration can be very effective in the presence of
shear turbulence that is naturally established in jet entrainment of
the surrounding ambient material.  It is also uncertain whether such
kinematic boosts to particle energies in either of these extragalactic
environs can enhance the sources' ability to generate cosmic rays with
\teq{E\gtrsim 10^{19}}eV, an issue that should be the focus of future
research.

\section*{Ultra-High Energy Cosmic Rays}
 \label{sec:uhecrs}

The study of Ultra-High Energy Cosmic Rays (UHECRs) bridges the
interests of cosmic ray physicists and astrophysicists.  While the
perennial problem of what is the metallicity of \teq{> 10^{19}}eV
cosmic rays (i.e.  protons vs. Fe) remains, focus at this meeting was
centered on the highest energy ones, namely those around and above the
Greisen-Zatsepin-Kuzmin (GZK) cutoff at \teq{\approx 5\times 10^{19}}eV
\cite{Greisen66,ZatKuz66}.  This subject was driven largely by the
recent announcement (Takeda et al. \cite{Takeda98}) that there is a
significant excess of cosmic rays above the GZK cutoff, with 13
events now detected (mostly AGASA data) above \teq{10^{20}}eV.  Papers at
the meeting can be categorized as those discussing arrival directions
and those addressing spectral issues.

Stanev and Hillas (OG 3.3.04) provided a detailed statistical analysis
of arrival directions for events with energies \teq{E>40}EeV, exploring
possible associations and anisotropies on various angular scales.
Their conclusions were that there is no significant correlation between
UHECR directions and those of extragalactic supernovae, and that there
was only a marginal enhancement of UHECR flux near the supergalactic
plane.  Ion deflections in galactic and extragalactic magnetic fields
clearly de-correlate directions of prospective sources and observed
events significantly.  Tkaczyk (OG 3.1.14) posited upper limits to the
neutron content of UHECRs via analysis of their anisotropy, using the
fact that neutrons are undeflected by these magnetic fields.  Stanev
and Hillas did indicate, however, that there was significant clustering
on angular scales less than 5$^\circ$, primarily spawned by two UHECR
triplets; pair groupings were not unusually numerous.  The Auger
\cite{Pryke98} and Owl \cite{Streit98} projects will obviously increase
the database dramatically, and improve such statistical analyses
immeasurably.  Directional information was also a focus of Horns et
al.  (OG 3.2.24), who used data from the HEGRA scintillation array to
search for high-energy gamma-ray associations with UHECR events, and
concomitant anisotropies.  One particular marginal association stood
out, a \teq{4\sigma} excess in the sky at gamma-ray energy of
\teq{10^{14}}eV, coincident with the arrival direction of the 320 EeV
Fly's Eye cosmic ray.  In a paper supporting this directional analysis,
Horns (OG 3.2.37) simulated electromagnetic cascades initiated by
UHECRs.

Two discussions relating to extragalactic source spatial distributions
were offered by Ptuskin, Rogovaya \& Zirakashvili (OG 3.2.23) and
Medina-Tanco (OG 3.2.52).  These two works focused on explaining the
excess implied by the UHECR observations \cite{Takeda98}, with
essentially the same premise: natural clustering of galaxies provides
source densities that exceed, on small distance scales, the average
density for a uniform, homogeneous spatial distribution.  This property
obviously weights the calculation of cosmic ray cooling by photo-pion
production on the microwave background, and permits a population of
UHECRs above the traditional GZK cutoff at \teq{\approx 5\times
10^{19}}eV.  Both groups effectively assumed that cosmic ray production
rates trace galaxy luminosity to some extent, since the latter
underpins astronomical detectability.  Ptuskin et al. and Medina-Tanco
reached the same conclusion: that the galaxy distributions can permit
cosmic ray distributions commensurate with the observed spectrum,
thereby resolving any purported observation/theory discrepancy.  Their
conclusion was arrived at by different analyses:  Ptuskin et al.
invoked a fractal distribution of galaxies as a
mathematically-motivated description of clustering, while Medina-Tanco
made use of the data collection of the CfA survey at redshifts
\teq{z<0.05}.  Hence the bottom line here is that there appears to be
no need to seek a galactic connection for the \teq{>10^{20}}eV events.

Papers addressing the actual source of UHECRs were exceedingly sparse,
with the only offerings being the galactic scenarios of Olinto, Epstein
\& Blasi (OG 3.3.03) and Blasi (OG 3.3.02).  Olinto et al. envisage
neutron stars acting as sources of ultra-high energy Fe, stripped off
the stellar surfaces by intense electric fields induced by rotation.
Key properties of their picture include a very flat source spectrum,
modelling structure around and above the ankle in the cosmic ray
spectrum, and of course, a heavy metallicity of the UHECR population.
Conditions for minimal effects of energy degradation of accelerated Fe
nuclei on the surrounding pre-supernova ejecta are achieved for fast
rotators, i.e.  millisecond pulsars.  Their model has a number of
attractive features, however its viability is contingent upon the ease
with which iron can be stripped from the star; this issue is somewhat
controversial in the pulsar community, with skeptics (in the majority)
appealing to the large work function of Fe to argue their case.  Blasi
(OG 3.3.02) suggested an exotic origin: super-heavy dark matter in the
galactic halo, comprising postulated quasi-stable particles that are
relics of the early universe.  These particles are purported to spawn
neutral and charged pions in spontaneous decays so that electromagnetic
signatures are generated, principally gamma-rays in the \teq{>100} MeV
range appropriate for exploration by the proposed GLAST
\cite{Gehrels99} experiment.  This scenario suffers from the drawback
that it is difficult to discriminate spectrally its predictions from
those of more mainstream origins of diffuse emission.  In a related
paper, Medina-Tanco \& Watson (OG 3.1.17) indicated that present
statistical limitations on UHECR anisotropies preclude discrimination
between various dark matter halo distributions.  Due to the proximity
of their sources, neither of these origin scenarios need to address
so-called GZK-violations.

\section*{Future Directions}
 \label{sec:future}

To conclude, it is appropriate to identify a list of salient tasks for
the cosmic ray community relating to the subjects discussed here.  For
origin/composition specialists, the question of how old the seed
material is still remains, and a reconciliation of ACE data constraints
with inferences from the LiBeB problem is needed.  Data on actinide
abundances should help probe matter mixing timescales.  It is also
important to determine whether O metallicity is a better indicator than
Fe/H for the LiBeB problem.  For the propagation community, extending
the data range of unstable secondary to primary ratios to span the
trans-relativistic regime, 1--10 GeV/nucleon, will help discriminate
between propagation models; while ACE has made progress here, we await
future flights of ISOMAX.  Improving spectra and composition studies
around the knee are clearly a major priority for the acceleration
community, to discern how effective SNRs are at accelerating up to
these energies.  A related issue is the search for pion decay
signatures in gamma-ray emission from remnants, which would provide the
first unequivocal proof that SNRs are indeed the galactic sites of
acceleration; the opportunity for this resolution seems imminent.  On
the theoretical side, three-dimensional plasma simulations are
desperately needed to elucidate the electron injection problem, and
considerable investment in the study of acceleration at relativistic
shocks would advance the astrophysics of active galaxies and gamma-ray
bursts.  For the UHECR field, it is anticipated that the database
increase due to Auger and Owl projects will provide a clearer picture
of the spectral, anisotropy and clustering properties of such high
energy particles, enabling discrimination between various postulates of
their origin.

\vskip 5pt\noindent
{\bf Acknowledgments:}  I thank my collaborators Don Ellison and Frank
Jones, and also Luke Drury and Bob Streitmatter for many insightful
discussions, and also for their critical reading of the manuscript.   I
also thank the Organizing Committee of the Conference for sponsorship
during my stay in Salt Lake City.

\end{document}